\documentclass[useAMS,usenatbib]{mn2e}   % mn2e use European A4, not suitable for arxiv
\usepackage{graphicx}
\title[R-T instabilities in SNRs Ia undergoing CR]{Rayleigh-Taylor instabilities in type Ia supernova remnants undergoing cosmic-ray particle acceleration 
       \\ - low adiabatic index solutions}

    \author[Chih-Yueh Wang]{Chih-Yueh Wang$^{1}$\thanks{E-mail: cywang@phys.cycu.edu.tw; cw5b@msn.com}
     \thanks{}\\
     $^{1}$Department of Physics, Chung-Yuan Christian University, 200 Chung-Pei Road,
    Chung-Li 32023, Taiwan}

\begin{document}

%\date{Accepted 1988 December 15. Received 1988 December 14; in original form 1988 October 11}

\date{}

\pagerange{\pageref{firstpage}--\pageref{lastpage}} \pubyear{2010}

\maketitle

\label{firstpage}

\begin{abstract}

This study investigates the evolution of Rayleigh-Taylor (R-T) instabilities in Type Ia supernova remnants that are associated with a low adiabatic index $\gamma$, where $\gamma<5/3$, which reflects the expected change in the supernova shock structure as a result of cosmic-ray particle acceleration.  
Extreme cases, such as the case with the maximum compression ratio that corresponds to $\gamma=1.1$, are examined.
%WR Extreme cases including an maximum compression ratio corresponding to $\gamma=1/1$ is examined. 
As $\gamma$ decreases, the shock compression ratio rises, and an increasingly narrow intershock region with a more pronounced initial mixture of R-T unstable gas is produced. Consequently, the remnant outline may be perturbed by small-amplitude, small-wavelength bumps. 
However, as the instability decays over time, the extent of convective mixing in terms of the ratio of the radius of the R-T fingers to the blast wave does not strongly depend on the value of $\gamma$ for $\gamma \ge 1.2$. As a result of the age of the remnant, the unstable gas cannot extend sufficiently far to form metal-enriched filaments of ejecta material close to the periphery of Tycho's supernova remnant. The consistency of the dynamic properties of Tycho's remnant with the adiabatic model $\gamma=5/3$ reveals that the injection of cosmic rays is too weak to alter the shock structure. 
Even with very efficient acceleration of cosmic rays at the shock,
%Even if the efficient acceleration of cosmic rays at the shock suffices, 
significantly enhanced mixing is not expected in Type Ia supernova remnants.
\end{abstract}

%\begin{online material}
 %movie animations

\section{Introduction}

Convective or Rayleigh-Taylor (R-T) instabilities occur in the intershock region of young supernova remnants (SNRs), whose evolution is dominated by the structure of the ejecta of the supernova SN. 

After the SN explodes, its fast ejecta are continuously decelerated by ambient material (interstellar or circumstellar medium).  Because the flow is supersonic, a blast wave propagates ahead of the spherical front of the expanding gas, or contact discontinuity (C.D.), whose pattern speed is equal to the flow speed of the ejecta.
%Since the pattern speed of the front, known as contact discontinuity (C.D.), is equal to the flow speed of the ejecta, the ejecta gas gradually piles up at the C.D. 
As the ejecta gas gradually piles up at the C.D., 
a high pressure is established and an internal shock %, or reverse shock, 
arises, which moves inward relative to the expanding ejecta in a Lagrangian sense. An SNR thus comprises two shells of heated, shocked gas, bounded by a forward and reverse shock, separated by a C.D. The reverse shock eventually turns back into the explosion center and terminates the free expansion of the supernova ejecta. In the intershock region, 
%since the inner dense shell of the shocked ejecta experiences outwardly-directed effective gravity, 
since the gas experiences outwardly-directed effective gravity, owing to deceleration, the flow becomes convectively unstable, such that the gas in the inner shocked shell protrudes into, and mixes with, the gas in the outer shell. Such an instability is only expected during an early SNR phase of expansion, when the SN ejecta bear a sufficiently sharp density distribution relative to the ambient medium, enabling the convection to persist 
(Wang \& Chevalier 2001 \& 2002; hereafter WC01 \& WC02).

%%%%%%%%%5

%%%%%%%%

Rayleigh-Taylor instability has been considered to be effective 
for modifying the spherical morphology of an SNR.  
A commonly observed SNR with irregular remnant outlines is Cassiopeia A SNR (Cas A), a prototype of Type II SNRs, of which both optical (Fesen \& Gunderson 1996) and X-ray (Hughes {\it et al.} 2000)
observations reveal numerous bumps and protrusions on its roughly-spherical outline.  
SN 1986 J and SN 1993 J are other, extreme examples of core-collapse SNe (Bartel {\it et al.} 1991; Bietenholz, Bartel, \& Rupen 2001).
Such an irregular morphology has been explained by the strong intershock convection that spreads the ejecta material out and beyond the nominal radius of the forward shock,
but the extensively mixed structure, including the dispersal of $^{56}Ni$ to high velocities, is now believed to be imprinted by the R-T instability that occurs during an asymmetric explosion
 (Kifonidis {\it et al.} 2000, 2003; Hungerford, Fryer, \& Warren 2003), 
 although processes which occur after the explosion such as the nickel bubble effect, 
  in which the radioactive progenitors of the Fe 
  expand relative to their surroundings,     
 explaining the large filling factor of the observed Fe 
(Woosley 1988; Li et al. 1993),
 contribute to the mixing. 
With respect to young Type Ia SNRs,
X-ray observations of Tycho's remnant (SN1572) 
reveal heavy ejecta material and dense knots, which, in non-decelerating motion, reach close to, and protrude from, the eastern periphery of the remnant 
 (Hughes, 1997; Hwang, Hughes, \& Petre 1998).
 Radio observations of Tycho's remnant (Reynoso {\it et al.} 1997) similarly revealed that the northeastern and the adjacent southeastern parts protrude from the circular outline; the southeastern part corresponds to the X-ray knots, 
while a regularly spaced structure immediately beneath the sharply marked forward shock front is present at the northeaster edge 
(Velazquez {\it et al.} 1998). 
 The large extension and non-decelerated motion of these structures are attributed to dense clumps in the diffuse supernova ejecta that expand into the remnant shell, rather than the global R-T convection that is caused by deceleration of the remnant (WC01).

    Although various hydrodynamic simulation models have been applied to study R-T instability in SNRs, most use a power-law density distribution of the SN ejecta, which yields a self-similar solution for the shock structure and applies only to the early expansion phase of core-collapse SNRs, when the inner flat component of the ejecta has not expanded so far as to change the intershock structure 
(Chevalier, Blondin, \& Emmering {\it et al.} 1992; Jun \& Norman 1996a; Blondin \& Ellison 2001, hereafter BE01, WC02; Fraschetti, 2010) 
Moreover, these idealized models with an adiabatic index of $\gamma=5/3$, such as that of Chevalier {\it et al.} (1992), have revealed rather limited R-T mixing. Jun \& Norman (1996a and 1996b) reached a similar conclusion after considering rapid cooling (Chevalier \& Blondin 1995) in the shocked ejecta. 
Jun \& Jones (1999) used accelerated cosmic-ray electrons as test particles to calculate the radio synchrotron emission but emphasized the reverse shock. Since rapid cooling and efficient cosmic-ray particle acceleration (Ellison, Berezkho, \& Baring 2000) both enhance the shock compression of the gas, BE01 adopted a low adiabatic index $\gamma<5/3$ to examine the instability, which exhibited the change expected because of particle acceleration. In this case, the mixing extends far enough to reach the forward shock front and considerably perturb the remnant outline. 
Recently, 
Fraschetti {\it et al.} (2010) employed $\gamma=4/3$ to represent the relativistic cosmic-ray particles and Ferrand {\it et al.} (2010) applied a kinetic model of nonlinear diffusive particle acceleration to the same hydrodynamic model as in Fraschetti {\it et al.}; 
%allowing a time-dependent back-reaction of cosmic rays with the shock;
they found that the development of instability is not strongly enhanced, independently of whether the cosmic rays dominate the reverse shock, or the shocks receive a back-reaction of cosmic rays that evolves with time. 
%
%Recently, Fraschetti  et al. (2010) used $\gamma=4/3$ at the forward shock to 
%represent the relativistic cosmic-ray particles and found that the development of instability is not strongly enhanced, 
%%whether the reverse shock is dominated by cosmic rays or not OR 
%independently of whether the reverse shock is dominated by cosmic rays. 
%Ferrand  et al. (2010) 
%%incorporated a kinetic model of nonlinear diffusive particle acceleration into the same hydrodynamic model OR 
%applied a kinetic model of nonlinear diffusive particle acceleration to the same hydrodynamic model, 
%allowing a time-dependent back-reaction of cosmic rays with the shock; 
%a more compact shock structure is revealed, but the R-T instability remains not %significantly affected.

Notably, the structure used in the above hydrodynamic models is not appropriate for Type Ia supernovae (SNe Ia), whose density falls exponentially with increasing velocity and is significantly different from that given by the power-law model (Dwarkadas \& Chevalier 1998; Dwarkadas 2000; WC01). 
An exponential profile is formed probably because the subsonic explosion of SNe Ia steadily releases energy behind a burning front, unlike a pure shock acceleration in core-collapse SNe.
The greatest dynamical contrast between the two models is that, because the exponential ejecta density gradient decreases with expanding SNR, the convection in Type Ia SNRs decays over time. As a result of the age of the remnant, the mixture of metal-enriched gas near the blast wave of Tycho's remnant cannot be attributable to an intershock convection. 
%{\sf rather, 
%it is best explained by 
%the expansion of a spherical shell of dense clumps in the supernova ejecta that expand into the remnant shell (WC01).
%} %\sf
Based on the indication that ejecta clumps are broken-up components of the shell of an inflated Ni bubble in the center of supernova ejecta that is heated by radiation from the $^{56}Ni\rightarrow^{56}Co\rightarrow^{56}Fe$ decay sequence 
(Basko 1994, WC01), the author (Wang 2005 \& 2008) utilized radiative transport radiation-hydrodynamic simulations to examine the structure and expansion properties of the inferred clumps in core-collapse and Type Ia SNe. For Type Ia SNe, when the density distributions produced by successful, W7-like SN explosion conditions are applied to the ejecta substrate, the properties of the inferred clumps is maximally compatible with those of the X-ray knots and filaments near the outline of Tycho's remnant (Wang 2008). 
In contrast, Warren {\it et al.} (2005) argued that, given the {\it Chandra} X-ray evidence of efficient cosmic-ray particle acceleration in Tycho's blast wave, the observed extended mixture of ejecta material could be explained by
 the R-T convection in the remnant shell, as was proposed by BE01. These investigations cast doubt on whether the enhanced intershock instabilities, expected to result from cosmic-ray particle acceleration, are responsible for the mixing present at Tycho's periphery.

This study examines the R-T instabilities in Type Ia SNRs using a low adiabatic index to approximate the expected increase in shock compression ratio caused by cosmic-ray particle acceleration. This study differs from the previous work of BE01 in that it adopts an exponential ejecta model to describe the general structure of Type Ia SNe, allowing the SNR evolutionary state to change with time. The rest of this paper is organized as follows. Section 2 introduces the effect of particle acceleration on SNRs. Section 3 describes the computational setup and methods. Section 4 then elucidates the evolutions of the R-T instabilities in one and two dimensions. Finally, section 5 and 6 presents discussions and draws conclusions.

\section[]{Acceleration of Cosmic-Ray Particles and Associated Shock Compression Ratio}

%\section[]{Cosmic-Ray Particle Acceleration and Shock Compression Ratio}

	Cosmic rays (CR) are electrons and atomic nuclei, mostly protons, that move at nearly the speed of light. Their energies can exceed $10^{21} \rm{eV}$. Strong shock waves of supernova remnants are believed to be efficient accelerators and primary sources of Galactic cosmic-ray particles, and to boost the energies of such particles to the knee of the CR spectrum, $10^{15} \rm{eV}$ (Axford 1981). The signature of this process is the decay of pions, which are generated in collisions of accelerated protons with atom and molecules  %interstellar matter, 
%while the shocks collide with the atom and molecules in the surrounding ISM.
while the shocks pass through the surrounding interstellar medium (ISM).
Electrons are also known to be accelerated to cosmic-ray energies in supernova remnants, 
via the scattering of low-energy photons, mostly the 2.7K cosmic background photons, which become TeV gamma rays through inverse Compton scattering (Koyama {\it et al.} 1995; Mastichiadis \& de Jager 1996, and Tanimori {\it et al.} 1998 for observations of SN 1006).
To explain the spectrum of SN 1006, Chevalier \& Reynolds (1981) first suggested that synchrotron emission, rather than other thermal and non-thermal processes, was responsible for the high-energy photons in SNRs. 
The energies of CR that are observed on Earth and the strength of the interstellar magnetic fields deduced from radio measurements support the prediction that the synchrotron radiation of CRs has X-ray wavelengths. 
Observations of X-ray synchrotron radiation from electrons ever since have shown that SNRs produce GeV emissions, although the synchrotron radiation in radio and optical wavelengths is not linked with the high-energy CRs (for a review see Reynolds 2008). 

	Although growing evidence relates cosmic-ray ions to the thermal properties of the shock-heated X-ray-emitting gas (Decourchelle, Ellison, \& Ballet 2000; Ellison, Slane, \& Gaensler 2001; Ellison {\it et al.} 2010), 
%{\it evidence of ion acceleration in SNRs remains circumstantial because synchrotron radiation in the X-ray spectrum is not easy to identify.}
 evidence of ion/proton acceleration in SNRs remains circumstantial. 
Cascade showers of optical photons from pion decay 
in RX J 1713.7-3946, 
which appear to be similar to SN 1006, 
%(and IC 273 ???),
even reveal the presence of TeV $\gamma$-rays (Enmoto {\it et al.} 2002).
The non-thermal X-ray emission from the shells of Galactic supernova remnants, most notably SN 1006 and Cas A, and
 the dominant non-thermal component in a few newly-discovered SNRs, believed to be the synchrotron emission from electrons that is accelerated to
$\sim 100$ TeV, represents evidence that cosmic rays are produced in SNR shocks (Petre, Hwang, \& Allen 2001).
However, 
most other remnants appear to be dominated by thermal emission from their shells, 
and disagreement exists about whether the $\gamma$-ray emissions in the direction of RX J 1713.7-3946 and others 
come from ions (Berezkho \& Volk 2006) or electrons (Katz \& Waxman 2008).
Reynolds \& Keohane (1999) calculated the maximum energies of shock-accelerated electrons in 14 young shell-type SNRs 
%assuming a constant magnetic field strength of 10 μG, 
and concluded that young remnants could not be responsible for the highest-energy Galactic cosmic rays.
A TeV proton has a $10^{-10}$ times lower critical synchrotron emission frequency than a TeV electron; however, identifying synchrotron emission from cosmic-ray protons in the radio wavelength is impossible, considering that a TeV proton produces a $\sim 10^{-3}$ times smaller flux than a TeV electron and many more GeV electrons than TeV protons contribute to the emission.  
Even if a spectral signature of proton/ion acceleration could be found, for which the best and only direct evidence relies on the gamma-ray emission that is not associated with inverse Compton emission from electrons while the indirect evidence comprises measurements of abnormally low electron temperatures and curvature in the SNR spectrum (Glenn Allen 2011, private communication; Allen, Houck, \& Sturner 2008), 
the evidence in support of the claim that isolated SNRs are the main accelerators of galactic CRs may be insufficient (Butt 2009).

%
%I think that the best way and perhaps the only way to provide direct
%observational evidence of the acceleration of protons and other ions in SNRs
%is to find evidence of gamma-ray emission that cannot be due to inverse
%Compton emission from electrons.
%
%Aside from the work of Warren et al., there are other indirect methods
%that have been used to infer the presence of cosmic-ray protons.  These
%include measurements of abnormally low electron temperatures and evidence of
%curvature in the shape of the electron spectrum of a supernova remnant.  For
%a (slightly dated) summary of this evidence, see Allen et al. 2008 (ApJ,
%683, 773).
%
%

To generate high-energy cosmic rays, the acceleration must transform at least 10\% of the total ejecta kinetic energy into relativistic particles, and this process cannot occur too soon ($\sim 100$ yr) after the supernova explosion, as in that case, the cosmic rays would lose most of their energy
(Blandford \& Eichler 1987; Drury, Markiewicz, \& Vo\"lk 1989; Dorfi 2000). 
% Given efficient particle acceleration, 
When efficient particle acceleration occurs, relativistic, superthermal particles can be created. The efficient production of superthermal particles is accompanied by decreased postshock temperature. Consequently, the region between the forward and reverse shocks of an SNR shrinks and becomes denser. This effect increases the compression ratio of the shock from $\sigma=4$, where
\begin{equation}
\sigma={{\gamma+1} \over {\gamma-1}}
\end{equation}
for strong shocks, to $\sigma>4$; and so the effective adiabatic index is less than 5/3 for the SNR intershock region. 
The increased compression of the gas gives rise to a more compact intershock structure and is expected to cause larger amplitudes for the R-T instability.

	In forecasting the gamma-ray flux from pion decay, Chevalier (1983) considered two-fluid, self-similar solutions where the ejecta was a thermal gas with $\gamma=5/3$ and the ambient medium was a relativistic gas that represented CRs with $\gamma=4/3$. In the case in which relativistic particles at the shock front contribute 100\% of the total pressure, 
%$\gamma_{eff}=1.333$, 
the effective adiabatic index is 1.333,
and the shock compression ratio is similar to that in the late radiative phase of SNR expansion, $\sigma>7$. 
BE01 showed that the results of the power-law model with $\gamma=1.1$ ($\sigma= 21$) are similar to the case of radiative cooling that was considered by Chevalier \& Blondin (1995) where the ejecta behaved as an ideal gas with gamma=1 and the ambient medium was modeled with $\gamma=5/3$.  
In kinetic simulations of diffusive shock acceleration (DSA), the cosmic-ray-modified shocks depend only weakly on the Mach number $M_0$ of the shock, and the total compression ratio is less than 10 even for $M_0 \sim 100$, because the propagation and dissipation of Alfven waves upstream reduces the CR acceleration and the precursor compression (Kang, Ryu, \& Jones, 2010). Based on these calculations, this work considers a maximum compression that corresponds to $\gamma=1.1$ for a young SNR such as Tycho's SNR, although theoretically arbitrarily large compression ratios are allowed. 	

	Berezhko \& Ellison (1999) demonstrated that the dependence of the compression ratio on the sonic and Alfven Mach number implies that the compression is most pronounced in SNRs with large shock speeds but relatively weak magnetic fields.
%, since the magnetic pressure counteracts the compression. 
However, for a shock with a high Mach number, $\sigma \sim 4$ can still hold if the injection of cosmic-ray particles into the shock is weak.

\section[]{Models and Methods}

% The computational methods used in this investigation follow those adopted in the author's previous work WC01.
  The two-dimensional hydrodynamic simulations performed in this investigation is based on the 
 %three-dimensional finite-difference parallel code ZEUSMP 
 %%a parallel version of ZEUS3D  
 the two-dimensional finite-difference code ZEUS2D
 %and its predecessor ZEUS2D 
 (Stone \& Norman 1992),
following the methods adopted in the author's previous work WC01.
 An exponentially-declining ejecta and a constant ambient interstellar medium 
is initialized on a spherical grid. 
 The flow velocity of the ejecta is freely-expanding and the ambient ISM is at rest.
 The inner boundary is inflow and fixed in radius, while the grid is radially nonuniform and expanding, following the motion of the forward shock and allowing finer resolution in the intershock structure.
%The grid velocities are initially the forward shock velocity times the radial fraction of the grid zones, and in every $\sim 100$ time steps the grid velocity are subtracted or added by an amount proportional to the shift in the radial fraction of the forward shock, such that the forward shock is kept at the same grid zone number.
A grid of 600 radial zones typically resolves the intershock region into $\sim 400$ zones.  % NW \gtrsim 400
Simulations usually begin as early as $10^6$ s after the supernova explosion, such that the flow can become numerically saturated and fully developed into the nonlinear regime long before Tycho's present epoch. 
Simulations initiated earlier presented numerical difficulty because the higher density reduced the numerical time step determined by the Courant condition.
The pressure distribution of the ejecta 
satisfies the adiabatic law $p=\kappa\rho^\gamma$, 
where $\kappa=6.1\times10^{14}$ (cgs units), %for $\gamma=4/3$, 
if the thermal energy is provided mainly by the entropy change during the nuclear burning of C and/or O to $^{56}Ni$ 
and $\gamma=4/3$ (Wang 2005).
%if the change in entropy during nuclear burning of C and/or O to $^{56}Ni$ provides most of the thermal energy
In the ejecta-dominated phase, the Mach numbers of the forward and reverse shocks greatly exceed 1, and so
the gas pressure used in the ejecta and ambient medium was trivial.

  The exponential density profile of the ejecta is characterized by an explosion mass $M$ and energy $E$, described by
\begin{equation}
        \rho_{SN} = A \exp(-v/v_e) \  t^{-3},
\end{equation}
where $v=r/t$ is the flow velocity of the ejecta, 
while $A$ is a constant, %$A={6^{3/2}\over 8\pi}{M^{5/2} \over M^{3/2}}$, 
$A= 6^{3/2} M^{5/2}/8\pi M^{3/2}$ , 
and $v_e$ is another constant called the velocity scale height, 
%$v_e=({E\over M})^{1/2}$;
 $v_e=({E/M})^{1/2}$;
 both constants are determined by the total mass $M$ and kinetic energy $E$ of the ejecta, derived by integrating the mass density and the kinetic energy density over space.
%$A={6^{3/2}\over 8\pi}{M^{5/2} \over M^{3/2}$ and $v_e=({E\over M})^{1/2}$.
 %, i.e., $M=8\pi A v_e^3$ and $E=48\pi A v_e^5$.
%The creation of an exponential profile for SNe Ia is probably related to the fact that the explosion energy is steadily released behind a burning front, unlike in the case of pure shock acceleration in core-collapse SNe. 
 A comparison with the power-law
 model yields the approximate power index of the exponential density,
 $n =  - d ln \rho/ d ln r =  v/v_{e} = r/v_{e}t$.
 Therefore, the exponential models with higher velocity scale (increasing with $E/M$ ratio) bear a flatter density
 distribution at a given flow velocity, and the density gradient increases with radius while it decreases over time.
Such an exponential profile closely approximates the outer density distribution of the free ejecta that is obtained from many successful explosion models (Dwarkadas \& Chevalier 1998), and 
the properties of the inferred ejecta clump 
in the most favorable explosion scenario, including the W7 deflagration model and the delayed detonation models, which depends on a near-Chandrasekhar-mass C-O white dwarf explosion, are compatible with those needed by Tycho's knots (Wang 2008).

% and the created clump properties in the most favored explosion scenario,
% i.e., the explosion of a near-Chandrasekhar-mass C-O white dwarf %that has accreted mass through Roche-lobe overflow, % from an evolved companion star,
%triggered by compressional heating near the white dwarf center,
%%including the deflagration model W7 and the delayed detonation models,
% are compatible with those needed by Tycho's knots (Wang 2008).

%Wang (2008) and Dwarkadas and Chevalier (1998) provide further details.

 This study used an explosion mass of $M_{1.4}$=1 and an explosion energy of $E_{51}$=1, where $M_{1.4}$ represents the mass in terms of the Chandrasekhar mass, 1.4$M_\odot$, and $E_{51}$ denotes the energy in units of $10^{51}$ ergs,
 along with a constant ambient density of $\rho_{am}$=$2.34\times10^{-24} \ \rm gm \ cm^{-3}$
(corresponding to a hydrogen number density $n_0=1 \ \rm{cm^{-3}}$ with a H/He ratio of 10/1 by number).
  Since the deceleration of an SNR is caused by the ambient medium,
 the interaction of the SN ejecta with a denser surrounding medium % prompts the remnant to evolve to a later phase.
  prompts a later evolutionary phase for the remnant.
 Nonetheless, solutions based on this set of
 parameters can be scaled into non-dimensional solutions,
 which can be re-scaled to obtain other dimensional solutions corresponding to
various sets of $M$, $E$ and $\rho_{am}$.
 One evolutionary sequence in the non-dimensional variables thus represents
 virtually all possible dimensional solutions.
 The following uses dimensionless variables, $r'=r/R$, $v'=v/V$ and $t'=t/T$, 
where 
$R'= ( {3M / 4 \pi \rho_{am}} )^{1\over 3}$,
$V'=( {2E / M} )^{1\over 2}$, and $T'= {R'/ V}$ are the scaling parameters,   
as described in WC01 to express the solution.
%The following use the scaling parameters as described in WC01: 
%$R'= ( {3M / 4 \pi \rho_{am}} )^{1\over 3}$,
%$V'=( {2E / M} )^{1\over 2}$, and $T'= {R'/ V}$, to describe the non-dimensional quantities $r'$, $t'$, and $v'$, 
%where $r'=r/R$, $t'=t/T$, and $v'=v/V$. 
For $M_{1.4}=1$ and $E_{51}=1$, $R=2.19 \ \rm pc$, $V=8.45\times 10^{3} \ \rm km s^{-1}$, and $T=248 \ \rm yr$. 
%
%for E_{51}=1.3$, V=9.63e3, T=217 yr.

% The following uses dimensionless variables $r'=r/R$, $v'=v/V$ and $t'=t/T$, %where $r'=r/R$, $t'=t/T$, and $v'=v/V$ 
%%as described in WC01 and Wang (2008) to express the solution. % $r$, $v$, and $t$. %results
%%The scaling parameters are defined as
%where
%\begin{equation}
%R'= ( {3M \over 4 \pi \rho_{am}} )^{1\over 3} \approx \ 2.19 \ {M_{1.4}}^{1\over 3} \ {n_0}^{-{1\over 3}} \ \ \rm{pc},
%\end{equation}
%\begin{equation}
%V'=( {2E \over M} )^{1\over 2} \approx \ 8.45\times 10^{3} \ ( {E_{51}\over M_{1.4}} )^{1\over 2} \ \ \rm {km \ s^{-1} },
%\end{equation}
%and
%\begin{equation}
%T'= {R'\over V} \approx \ 248 \ {E_{51}}^{-{1\over 2}} \ {M_{1.4}}^{5\over 6} \ {n_0}^{-{1\over 3}} \ \ \rm{yr},
%\end{equation}
%to express the solution. % $r$, $v$, and $t$. %results

\section[]{Evolution of Instabilities}
\subsection{One-dimensional Simulations}

\begin{figure}
% \vspace{302pt} 
 \centering
  \includegraphics[scale=0.45]{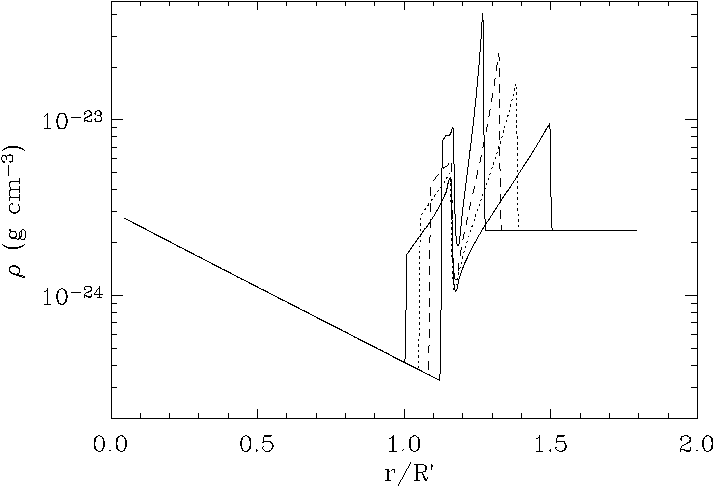} % pdflatex can use png/jpg OK
%\label{f01}
 \caption
 {
Radial profile of gas density in one-dimensional exponential model at dimensionless age $t'=1.75$ for $\gamma$=5/3, 4/3, 1.2, and 1.1, corresponding to Tycho's present state. Ratios of forward shock radius/reverse shock radius to reverse shock radius/forward shock radius are 1.51/0.66, 1.33/0.71, 1.22/0.81, and 1.15/0.87, respectively. 
 }
 \label{fig1}
 \end{figure}

% standard model (2010-1572)*365.24*24*3600.0/Tp=1.77374
% W7 model       (2010-1572)*365.24*24*3600.0/Tp=2.02235
One-dimensional hydrodynamic simulations of the cases $\gamma$=5/3, 4/3, 1.2, and 1.1 were performed first. Figure \ref{fig1} presents the one-dimensional density profile 
 in an age $t'=1.75$, 
   % using hdf083eU,jU,gU,fU, IDL> @tycho3d-1e8 time,timen 1.36001E+10  1.75811
which is approximately the normalized age of Tycho's remnant, 
 $t_0=1.77$ (438 years), if standard explosion parameters are adopted.
If a W7-like explosion condition, $M_{1.4}=1$ and $E_{51}=1.3$, is assumed,                                                               then Tycho's true age corresponds to a dimensionless age of $t_0=2.02$.  
%Figures \ref{fig2}, \ref{fig3}, and \ref{fig4} 
Figures 2, 3, and 4 
illustrate the density distributions in the case of $\gamma=4/3$, 1.2, and 1.1 at four various stages, overplotted with the angle-averaged radial profile.
For $\gamma=5/3$, the reverse shock begins to turn inward in the stellar frame at $t'=2.5$, and reaches the stellar center at $t'=8$. 
%{\it Because the shocked ISM and ejecta must have their pressures balanced at the contact discontinuity, in the non-adiabatic cases, the intershock layer is further compressed toward the contact discontinuity with the position of the contact discontinuity being unchanged, and the turnover of the reverse shock and its subsequent impact with the explosion center are significantly delayed. }
In the non-adiabatic cases, because the sound speeds of the hot, shocked ejecta and hot, shocked ISM are still high and permit no pressure difference at their interface, 
the pressure-equilibrated contact discontinuity remains fixed in space while
the intershock layer is further compressed toward the contact discontinuity.
Accordingly, the turnover of the reverse shock and its subsequent impact with the explosion center are significantly delayed.
X-ray observations have suggested that the reverse and forward shocks in Tycho's remnant are located at radii of 2.08 pc ($r'=0.98$ in dimensionless units) and 3.09 pc ($r'=1.45$), and decelerate with deceleration parameters of 0.15 and 0.47, respectively. (See references in Dwarkadas \& Chevalier 1998 and Warren {\it et al.} 2005). Notably, these observed position and deceleration of the shocks 
 are consistent with the result obtained using the exponential model with $\gamma=5/3$, 
 but differ greatly from those obtained with $\gamma=1.1$; 
%{\sf at $\gamma=5/3$, $\gamma=4/3$ and $\gamma=1.2$,
 at $\gamma=5/3$, 
 %the reverse to forward shock radius ratios are $\sim 2/3$, %0.75 and 0.81, respectively, 
 the ratio of reverse to forward shock radii is $\sim 2/3$, %0.75 and 0.81, respectively, 
 whereas $\gamma=1.1$ yields an extremely high value of 0.87.
%The use of small $\gamma$ causes the SNR to suffer from energy losses during expansion, and so 
% the deviation of the shock radius ratios from the observed values increases with time and decreasing $\gamma$.
The deviation of the shock radius ratios from the observed values increases with time because the use of small $\gamma$ causes the SNR to suffer from energy losses during expansion.

%
% FIG_prof1d_2010
%  gamma   FS/RS   RS/ FS
%   5/3    1.50649 0.0.663795
%   4/3    1.33219 0.750643
%   1.2    1.23485 0.809814
%   1.1    1.14511 0.872973

% ----------- small figure ----------
%\begin{figure}[!t]

% ----------- small figure ----------
\begin{figure}
\centering
  \includegraphics[scale=0.45]{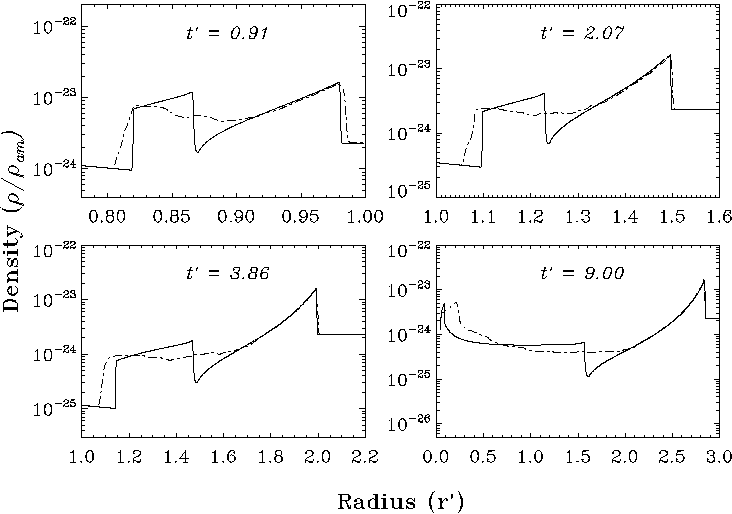} % pdflatex OK
%  \includegraphics[width=3.0in]{dens1d-gam13.eps}
%\vbox{
\caption
{
Angle-averaged two-dimensional density distribution plotted with the one-dimensional 
unperturbed solution at four 
stages of the evolution with $\gamma=4/3$. 
The time and radius use the normalized units given in the text.
Two-dimensional solutions apply a $20\%$ 
perturbation at the contact discontinuity.
% The density is normalized to the ambient density.
% In two dimensions, the reverse/forward shock smears to inner/outer radius owing to instabilities.
}
%} %vbox
\label{fig2}
\end{figure}

% ----------- small figure ----------
\begin{figure}
\centering
  \includegraphics[scale=0.45]{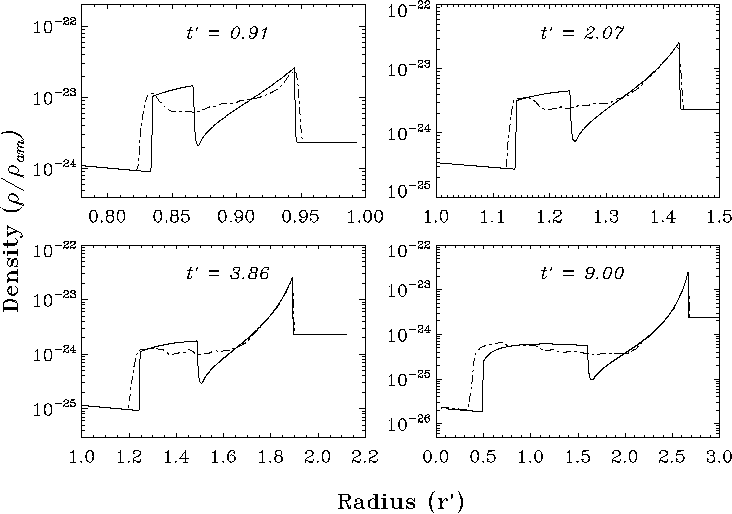} % pdflatex OK
\vbox{
\caption
{
Angle-averaged two-dimensional density distribution plotted 
with one-dimensional unperturbed solution at four stages of evolution with $\gamma=1.2$. 
Two-dimensional solutions apply a $20\%$ perturbation below the reverse shock.
}
}  % vbox
\label{fig3}
\end{figure}

% ----------- small figure ----------
\begin{figure}
\centering
 \includegraphics[scale=0.45]{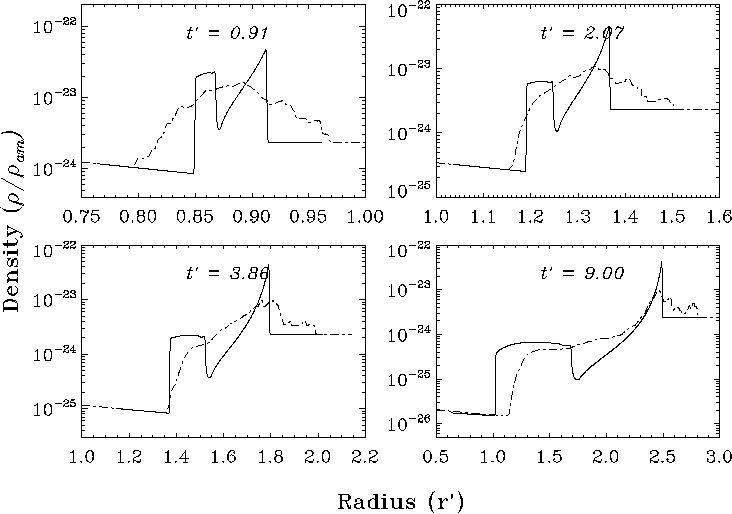} % pdflatex OK
% \includegraphics[width=3.0in]{dens1d-gam11.eps}
%\vbox{
\caption
{
Angle-averaged two-dimensional density distribution plotted with the one-dimensional
unperturbed solution at four
stages of the evolution with $\gamma=4/3$.
The time and radius use the normalized units given in the text.
Two-dimensional solutions apply a $20\%$
perturbation at the contact discontinuity.
The reverse and forward shocks are strongly perturbed 
and smear to an inner and outer radius respectively.
% owing to instabilities. 
}
\label{fig4}
\end{figure}

\subsection{Two-dimensional Simulations}

In most two-dimensional simulations, a density perturbation with an amplitude of 20\% across contact discontinuity 
%to produce the most efficient excitation, 
is applied to excite the instability as efficiently as possible.  
The instability was seeded early enough for the flow to evolve into the nonlinear regime
before Tycho's present epoch,
when the growth of R-T fingers is matched by their destruction due to shearing and advection.
%when %the rate of 
%the growth of R-T fingers is matched by their %rate of 
%destruction due to shearing and advection.
Figures 5 and 6
%NW Figures \ref{figgamma13} and \ref{figgamma12_1e8} 
plot the density contours for $\gamma=4/3$ and $\gamma=1.2$, respectively. 
Following an initial growth phase, the initial perturbation grows to form mushroom-shaped caps. Shrinking of the intershock layer and increasing in the density contrast across the C.D. generate narrower, denser and sharper R-T fingers. The flow then becomes independent of its initial conditions and enters a nonlinear phase, in which an increasingly larger instability mode is built up. The reverse shock front becomes corrugated and moves slightly inward, and the forward shock is not much perturbed.
The evolution of the flow appears similar to that seen in WC01.
In the case of $\gamma=1.2$, 
the R-T fingers can exert strong pressure and slightly disrupt the blast wave outline  
only in the initial stage of linear growth. %the bumpy structures smooth down over time. } 
However, the fingers at saturation do not come close to the blast wave, and so do not affect them; 
 once the dense, shocked ejecta is mixed close to the forward shock, it quickly drops behind. The forward growth of the caps is impeded by the drag of the flow. The relative motion of flows between a finger and its surroundings bends the stems and disrupts the flow. 
The mushroom cap eventually falls off to the side. The remaining filaments are then swept back to the C.D. 
In the later stages, Kelvin-Helmholtz instability takes over, creating vortex rings in the less dense regions that are left by the original mushroom caps. 
%Strong outward convection develops from the reverse shock when the reverse shock accelerates back toward the center. 
The vortex rings gradually come to dominate the spikes, and only the stems remain recognizable.

In the case of $\gamma=1.1$ (Fig. 7), %\ref{figgamma11},  
the R-T fingers quickly extend to the forward shock in the linear stage of instability development, pushing small regions ahead of the average shock radius and distorting the outer blast wave. 
The R-T fingers exert strong pressure and disrupt the blast wave outline. 
The extent of the distortion seem to depend on the wavelength of the perturbation, 
and also the time since the initial perturbation is applied. 
However, as in the case of $\gamma=1.2$, the vortex rings eventually come to dominate globally, 
and the bumpy structures on the remnant outline smooth down over time.

The radius of the R-T fingers is compared to the forward shock radius. 
The extension of the R-T fingers is greatest in the stage of linear instability growth. 
At $\gamma$=1.1, 
the fingers pass the nominal shock radius. 
%At $\gamma=1.2$, however, the R-T finger reaches only $\sim 87\%$ of the blast wave radius in Tycho's present epoch. The extent of mixing is virtually identical to that in the case of $\gamma=5/3$ considered by WC01. 
At $\gamma=4/3$ and $\gamma=1.2$ (Fig. \ref{figgamma12_1e8}, however, the R-T finger reaches only $\sim 87\%$ of the blast wave radius in Tycho's present epoch. 
The extent of mixing is virtually identical to that in the case of $\gamma=5/3$ considered by WC01. 
For $\gamma=1.2$ with efficient perturbations,
the R-T fingers reach $\sim 93\%$ of the blast wave radius  
(Fig. \ref{figgamma12_1e8}) and slightly perturb the shock
(Fig. \ref{figgamma12_1e6}),  
but the mixture is much less evident
 than the dense gas immediately beneath the forward shock front.

The amplitude, wavelength and initial age of the perturbation were varied. WC01 demonstrated that the instability in the nonlinear regime is insensitive to the initial strength of the perturbation.
%%% this is because --> because
%%% fully developed in Tycho's remnant --> fully developed in Tycho's present epoch
Dwarkadas (2000) used the VH-1 code to indicate that the instability that arose from numerical noise remained in the linear regime, because viscosity and diffusion tend to dampen small-wavelength instability modes, which outpace all other modes; but with a 2\% perturbation, his results were consistent with WC01. 
When the reverse shock accelerates back towards the center, distinct Rayleigh-Taylor fingers that protrude outwards from the reverse shock front are formed,
when the intershock instability is not fully developed,
whereas in the nonlinear case, the mode of such convection is not obvious (Fig. 8). %(Fig.\ref{figspurious}).
During of the work on WC01, the flow could be induced into a nonlinear phase without an initial perturbation, as long as the simulation was begun soon enough with sufficiently high resolution, when the intershock structure had a high density contrast and large expansion rate. Given that Tycho's remnant shell has propagated for numerous inflight times since the explosion, 
the nonlinear instability should have become fully developed in the remnant.

%---------wide figures----------

% \begin{figure}[!t]
  \begin{figure*}
  \centering
 \vbox to 220mm{
 \caption
%$t'=0.0128$ ($1\times 10^{8} \rm sec$). 
%$t'=0.0449$ ($2.5\times 10^{8} \rm sec$) 
%$log_{10} \rho \rm(g cm^{-3})$.
{
Series of density contours that depict time evolution of dynamical instability using $\gamma=3/4$.
Simulation was initiated at 
 $t'=0.00025$ ($1\times 10^{6} \rm sec$). 
Initial density perturbation 
was imposed at 
 $t'=0.0006$ ($4\times 10^{6} \rm sec$) 
to ejecta across the contact discontinuity,
with 20\% amplitude in the $l=100$ mode of the spherical harmonic function. 
The grid has 600 nonuniform radial
zones by 300 angular zones in one half of a quadrant, 
which resolved the intershock region 
into $\sim 400$ radial zones. 
Contours correspond to logarithmically scaled values between the 
lowest and the highest values sampled in the intershock region. 
Color bars represent base-10 logarithmic values of density.
}
 } % vbox
  \label{figgamma13}
  \vspace{-190mm}
%   \includegraphics{thisplot3bw.eps} % time0=1e8
%    \includegraphics{thisplot3b_gamma13_time01e6.png} % time0=1e6
%----------------gamma=4/3 several small figures START-------
   \includegraphics[width=2.2in]{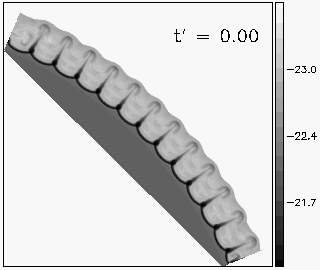} % pdflatex OK
   \includegraphics[width=2.2in]{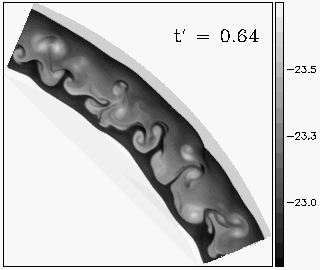} % pdflatex OK
   \includegraphics[width=2.2in]{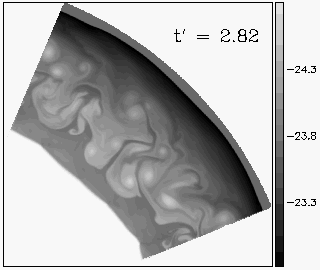} % pdflatex OK
\
   \includegraphics[width=2.2in]{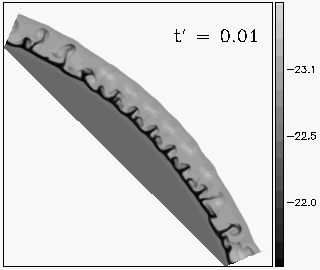} % pdflatex OK
   \includegraphics[width=2.2in]{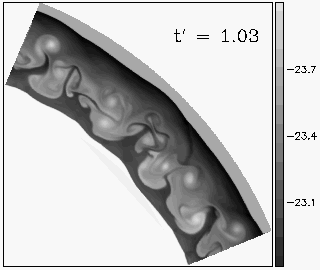} % pdflatex OK
   \includegraphics[width=2.2in]{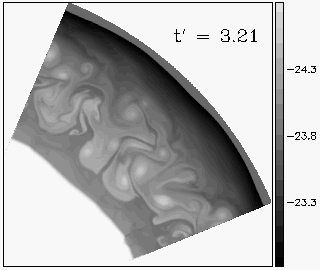} % pdflatex OK
\
   \includegraphics[width=2.2in]{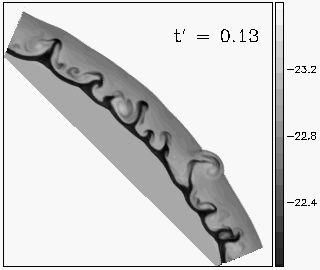} % pdflatex OK
   \includegraphics[width=2.2in]{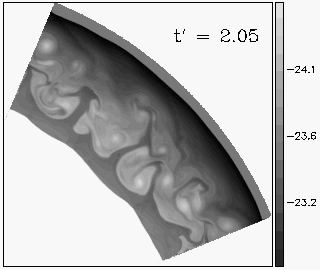} % pdflatex OK
   \includegraphics[width=2.2in]{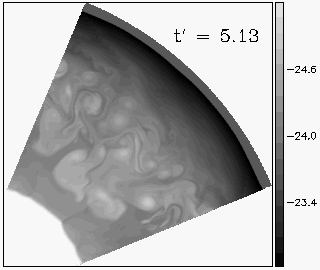} % pdflatex OK
%----------------gamma=4/3 several small figures END-------
%  \includegraphics[width=7.2in]{thisplot3bw.png} % pdflatex OK
 \end{figure*}

  \begin{figure}
  \centering
   \includegraphics[width=1.6in]{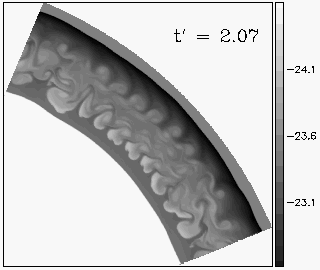} % BW time0=1e8, RS perturb 
   \includegraphics[width=1.6in]{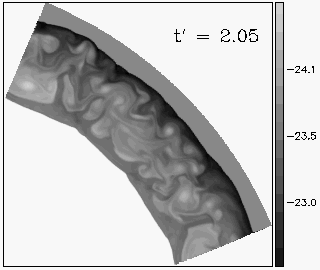} % BW time0=1e8, CD perturb
\
   \includegraphics[width=1.6in]{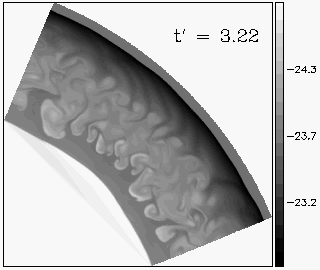} % BW time0=1e8, RS perturb
   \includegraphics[width=1.6in]{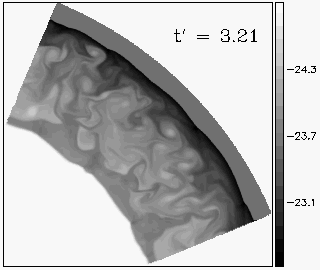} % BW time0=1e8, CD perturb
\
   \includegraphics[width=1.6in]{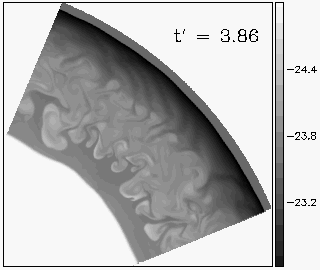} % BW time0=1e8, RS perturb
   \includegraphics[width=1.6in]{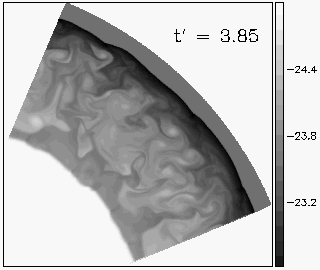} % BW time0=1e8, CD perturb
\
   \includegraphics[width=1.6in]{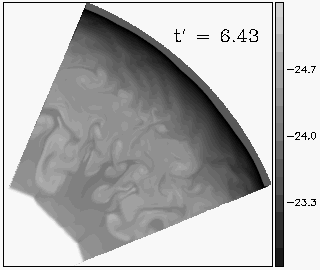} % BW time0=1e8, RS perturb
   \includegraphics[width=1.6in]{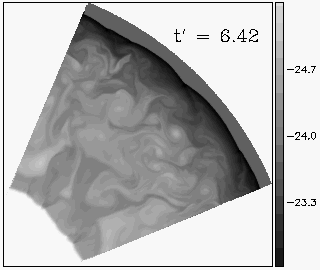} % BW time0=1e8, CD perturb
 \caption
{
Density images of the $\gamma=1.2$ model.
Left panel - 
Simulation was initiated at $t'=0.0128$ ($1\times 10^{8}$ s).
Initial density perturbation was imposed at $2.5 \times 10^{8}$ s to ejecta below 
reverse shock,
with $\sim 20\%$ amplitude and an $l=100$ mode of the spherical harmonic function.
Right panel - 
Simulation was initiated at $t'=0.00025$ ($1\times 10^{6}$ s).
Initial density perturbation was imposed at $4 \times 10^{6}$ s 
across contact discontinuity,
with $\sim 20\%$ amplitude and an $l=100$ mode of the spherical harmonic function.
The grid has 600 uniform radial zones by 300 angular zones in one half of a quadrant.
Contours correspond
to logarithmically scaled values between the lowest and the highest values sampled in 
the intershock region. Color bars represent base-10 logarithmic values of density.
%$log_{10} \rho \rm(g cm^{-3})$.
}
 \label{figgamma12_1e8}
 \end{figure}
% fig6

%-------gamma=1.2 normal figure in 1 column, 2x2, time0=1e6, perturb CD-------
% \begin{figure}[!t]
  \begin{figure}
  \centering
%  \includegraphics{ttx3y3_spacing_colbarv_full2_45_BW.eps}
%  \includegraphics{ttx3y3_spacing_colbarv_full2_45_BW.eps}
%  \includegraphics[width=7.2in]{ttx3y3_spacing_colbarv_full2_45_BW.png} % 
%   \includegraphics[width=1.5in]{hdf125gU33w.png} % 
%   \includegraphics[width=1.5in]{hdf180gU33w.png} % 
%   \includegraphics[width=1.5in]{hdf148gU33w.png} % 
%   \includegraphics[width=1.5in]{hdf200gU33w.png} % 
% fig7
   \includegraphics[width=1.5in]{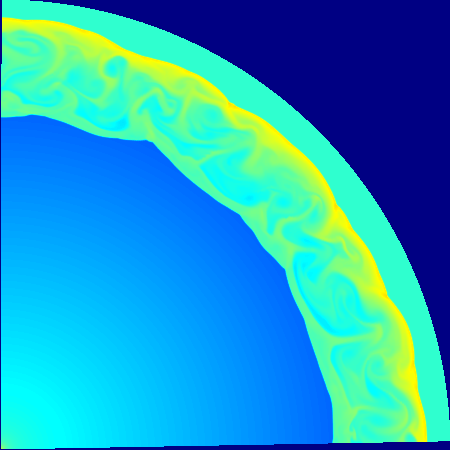} % blue and yellow, time0=1e6 
   \includegraphics[width=1.5in]{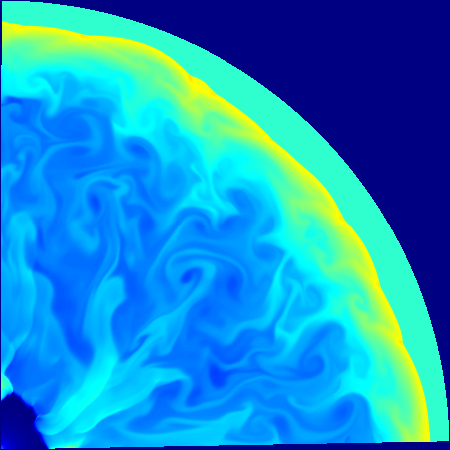} % 
   \includegraphics[width=1.5in]{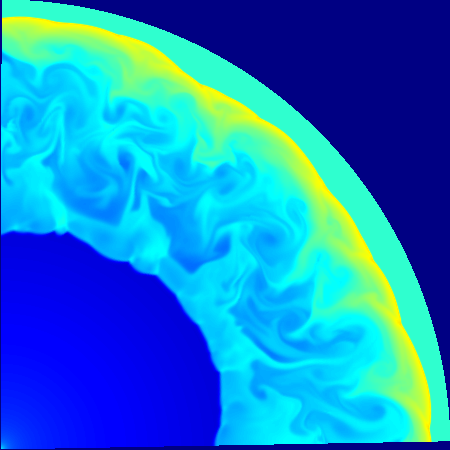} % 
   \includegraphics[width=1.5in]{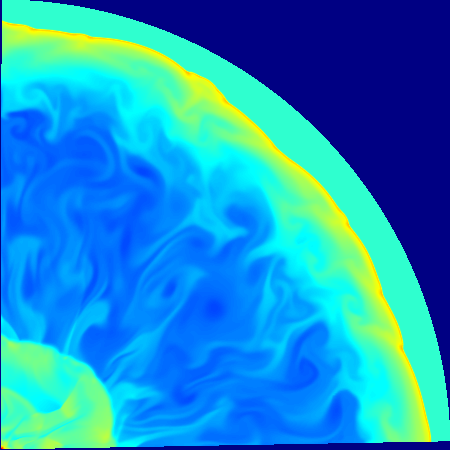} % 
 \caption
{
Density images of the $\gamma=1.2$ model at $t'=2.07$, 5.02, 9.12, and 11.69. 
Initial density perturbation was imposed 
at $4 \times 10^{6}$ s across contact discontinuity,
with 20\% amplitude and an $l=100$ mode of the spherical harmonic function. 
The grid has 600 uniform radial zones by 300 angular zones in one quadrant. 
}
 \label{figgamma12_1e6}
 \end{figure}
% fig7

% --------- start gamma=1.1 ---------
%\thispagestyle{plate}
%\plate{Opposite p. 812, MNRAS}
 \begin{figure*}
% \centering
% \includegraphics[width=2.5in]{f3.eps}
%NW  \includegraphics{ttx3y3_spacing_colbarv_full2_f4a-axis.eps}
%NW  \vspace{-300mm}
 \vbox to 220mm { %\vfil HELLLOW!!!!!!!!!
%Simulation was initiated at $t'=0.0128$ ($1\times 10^{8} \rm sec$). 
%$t'=0.0449$ ($2.5\times 10^{8} \rm sec$) 
%$log_{10} \rho \rm(g cm^{-3})$.
 \caption
{
Series of density contours that depict time evolution of 
dynamical instability using $\gamma=1.1$.
Simulation was initiated at $t'=0.00025$ ($1\times 10^{6}$ s). 
Initial density perturbation was
imposed at 
 $t'=0.0006$ ($4\times 10^{6}$ s) 
to ejecta below reverse shock, with 20\% 
amplitude in the $l=100$ mode of the spherical harmonic function. 
The grid has 600 nonuniform 
radial zones by 300 angular zones in one half of a quadrant, 
which resolved the intershock region into
$\sim 400$ radial zones. 
Contours correspond to logarithmically scaled values between the 
lowest and the highest values sampled in the intershock region. 
Color bars represent base-10 
logarithmic values of density.
}
           } % vbox
 \label{figgamma11}
  \vspace{-190mm}
%\includegraphics{ttx3y3_spacing_colbarv_full2_f4a-axis.eps}  % OLD
% \includegraphics{ttx3y3_spacing_colbarv_full2-f4a-axis-33.eps}
%  \includegraphics[width=7.2in]{ttx3y3_spacing_colbarv_full2-f4a-axis-33.png} % pdflatex OK
%----------------gamma=1.1 several small figures START-------
   \includegraphics[width=2.2in]{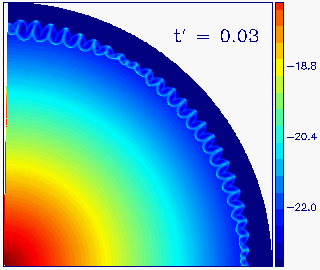} % pdflatex OK
   \includegraphics[width=2.2in]{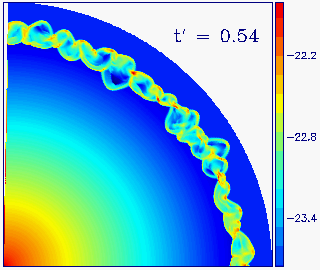} % pdflatex OK
   \includegraphics[width=2.2in]{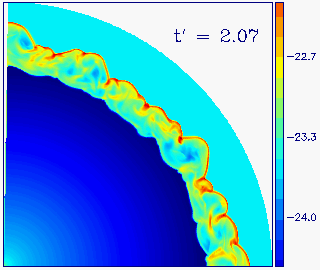} % pdflatex OK
\
   \includegraphics[width=2.2in]{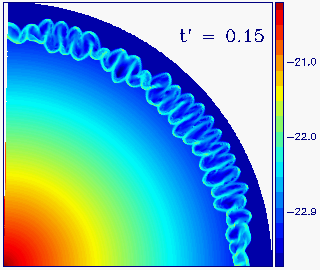} % pdflatex OK
   \includegraphics[width=2.2in]{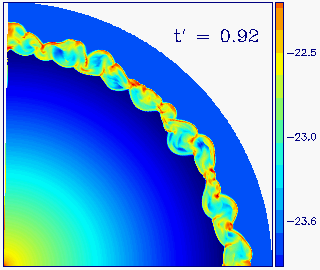} % pdflatex OK
   \includegraphics[width=2.2in]{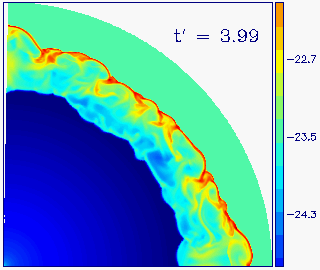} % pdflatex OK
\
   \includegraphics[width=2.2in]{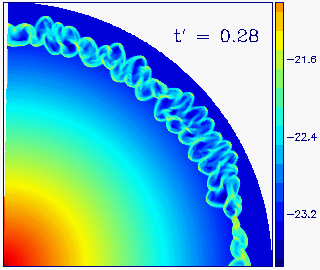} % pdflatex OK
   \includegraphics[width=2.2in]{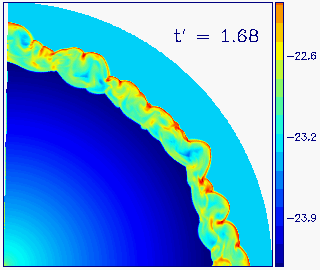} % pdflatex OK
   \includegraphics[width=2.2in]{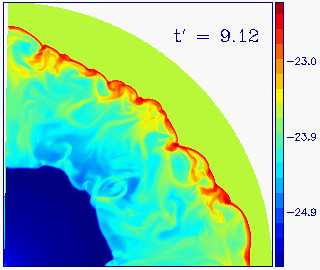} % pdflatex OK
%----------------gamma=1.1 several small figures END-------
 \end{figure*}
%fig8

%-----small figure---------
\begin{figure}   %[!t]
 \centering
%   \subfigure[]
   {
       \includegraphics[width=1.5in]{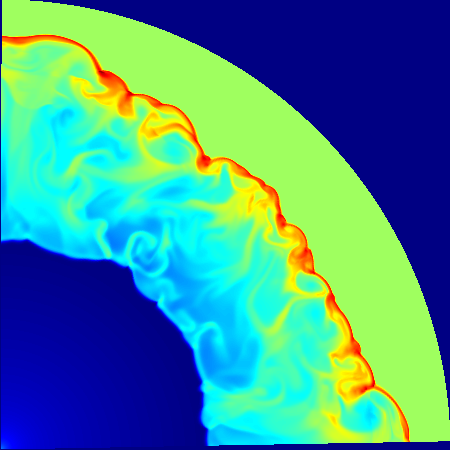}
%   \label{f01a}
    }
%  \subfigure[]
 {
     \includegraphics[width=1.5in,angle=0]{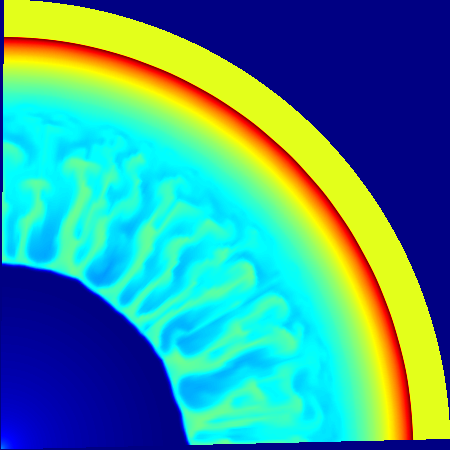}
  \label{f01b}
 }
% \label{f01b}
% fig9
\
 \caption
{
Density images of the $\gamma=1.1$ models that depict different results at $t'=7.83$, 
when the reverse shock has turned over. 
Contours correspond
to logarithmically scaled values between the lowest and the highest values sampled in
the intershock region. 
Left - Perturbation was imposed with 10\% amplitude and an $l=100$ mode of the spherical 
harmonic function. The grid has 300 radial by 600 angular zones.
Right - No perturbation was imposed. The grid has 600 radial by 100 angular zones.
The instability at the reverse shock displays a dominant mode because the instability in the intershock region
 is not fully grown. Such a result is spurious.
}
 \label{figspurious}
 \end{figure}

%-----small figure---------
 \begin{figure}
 \centering
   \includegraphics[scale=0.45]{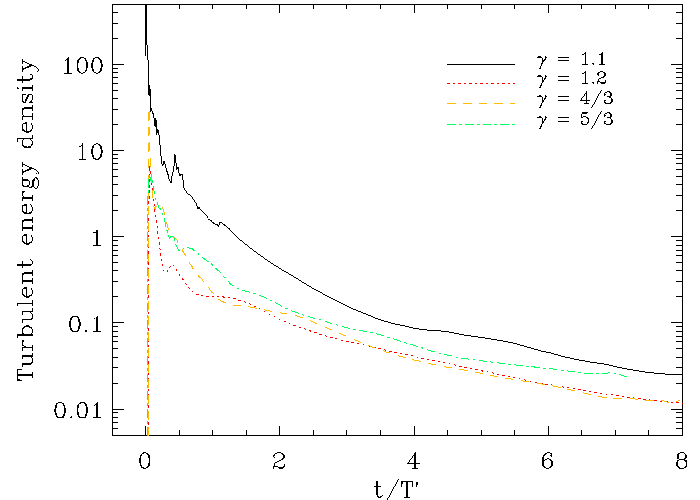} % pdflatex OK
 %\vbox{
 \caption
{
Evolution of angular turbulent energy density 
normalized to bulk mass in intershock region for 
four values of $\gamma$. 
Same angular resolution of 300 zones in one half of a quadrant is used
in all cases. 
Turbulence is measured in cgs units. 
Curve cross occurs when R-T fingers create bulges on blast wave.
}
 \label{figturb}
 \end{figure}
% fig10

%{\bf plot FS radius 2D/1D with time for gamma=1.1.} 

%% USE THIS to incluence the blast wave

The use of $\gamma=1.1$ easily generates sufficiently strong excitations to influence the blast wave, particularly in the early stage of an SNR.
The cases with 1\% perturbation initiated at $10^4$ s and with 20\% perturbation initiated at $10^6$ s both yield strong turbulence, such that the bulges on the remnant outline change rapidly in space and time.
The presence of the strong convection is not determined by the strength of the applied perturbation, but depends on whether there is a sufficiently large density contrast across the C.D.   
Notably, because the fingers reach the forward shock before saturation, the strong shearing flow that is generated while the blast wave is distorted blocks the growth of R-T fingers, which therefore cannot be pushed out beyond the shock, and their growth stalls. As shown in BE01, if a fragment (R-T finger) of dense ejecta has sufficient radial momentum to push out the forward shock in a local protrusion, then the deformed shock generates a substantial tangential post-shock shear flow around the ejecta finger, rapidly disrupting the finger or ejecta fragment through the Kelvin-Helmholtz instability, before it is advected back into the interaction region. 
Therefore, a local protrusion of the shock front of more than a few percent of the blast wave radius cannot be formed when the unstable flow has saturated into the nonlinear regime.

\section{Comparison between the Exponential model and the Power-law model}

The exponential model can be understood as a power-law model with a continuously decreasing power index $n$.
In BE01's power-law model, the mixture with the highest compression ratio ($\sigma=21$ and $\gamma$=1.1) quickly reaches and surpasses the average shock radius; the remnant outline are perturbed with low-amplitude, small-wavelength bumps, although the average shock front remains almost spherical. 
 Higher compression ratios is expected to cause significant deviation from spherical symmetry. 
In the exponential model, 
a preferred mode in increasing wavelength that is independent of the initial conditions develops toward the end of the evolution. 
The perturbation to the remnant outline is eventually dominated by the standard evolutionary paradigm that was illustrated in WC01 
and therefore depends on the age of the remnant. A strong perturbation that is seeded near Tycho's age may protrude out of the blast wave, but this phenomenon is caused by the fact that the instability has not grown into the nonlinear regime. 
Given that a fully-grown perturbation 
causes less distortion than the perturbation in the linear stage,
if the instability develops from a small perturbation soon after the SN explodes, 
then the remnant spherical outline should not be substantially altered.
The result is consistent with that in the dramatic case of $n=12$, $s=0$ (for a constant ambient medium, $\rho_{am}\propto r^{-s}$), and $\gamma=1.1$, considered by BE01, in which the deviation from a spherical outer shock is small. 
%Notably, the present density power index of the free ejecta at Tycho's reverse shock is only $n \approx 3$. 
Notably, the power index $n$ of the free ejecta underneath the reverse shock drops below 7 at $t'\ge 0.1$, and the present density power index at Tycho's reverse shock is only $n \approx 3$
(see Fig. 5 of WC01). In the power-law model self-similarity does not exist for $n \le 3$.

BE01 showed that, despite the large changes in the radial profiles, the width of the mixing region in terms of the ejecta mass fraction in the power-law model did not noticeably vary with $\gamma$.
The extent of mixing or the saturated instability growth rate, measured by summing the angular components of the turbulent energy density in the intershock region and then normalizing the sum to the bulk kinetic energy density that is associated with the forward shock front (to remove the expansion and compression effects in the radial direction), is also relatively unaffected by changes in $\gamma$. 
Hence, the dominant wavelength of the instability appears qualitatively similar  
for $\gamma=4/3$ and $\gamma=1.2$  
(see Fig. 4 in BE01, for $n=7$ and a circumstellar medium $s=2$).
In the exponential model, the maximum radius of the R-T fingers (in terms of the forward shock radius) does not vary significantly with $\gamma$. Furthermore, both $\gamma=4/3$ and $\gamma=1.2$ are associated with a similar mode (wavenumber of $\sim 4$)
to that in the adiabatic case of WC01 in Tycho's present epoch. The wavenumber is less than that in the adiabatic, power-law model with $n=7$ and $s=0$, as expected, where a half quadrant has six waves (Chevalier, Blondin, \& Emmering, 1992).
The instability has a delayed response to the change in the density gradient.
% similarity - resemblance
The similarity among the dominant instability modes at various $\gamma$ above 1.2
follows from the fact that the energy losses in a SNR cannot change the radius of the contact discontinuity 
and so considerably reduce the decelerations of the shocks in Tycho's present epoch. 
This result is consistent with that of BE01.
Notably, the case of $\gamma=1.2$ yield a 12\% lower speed of the blast wave than does $\gamma=5/3$,
whereas the case of $\gamma=1.1$ gives a 28\% lower shock speed, amounting to unrealistic losses in kinetic energy.

%{\bf 
%Figure 10
%%\ref{figveldecel} 
%plots the evolution of velocities and deceleration parameters of the forward shock for various $\gamma$. The case of $\gamma=1.1$ yields a 20\% lower speed and a 10\% smaller deceleration of the blast wave than does $\gamma=5/3$.
%} 

%{\sf  
% \begin{equation}\chi={\int\rho u_{\theta}^{2} d\tau  \over \rho_{1} V_1^{2} \int d\tau}
% \end{equation}
% where the integration is over the volume of shocked gas, 
%}

%---------------end of NEW -------

A major difference between the two models is that, in the power-law model, the instability is insensitive to changes in the compression ratio when the growth rate of R-T fingers (curves in Fig. 5 of BE01) become approximately flat, as when the system is in the nonlinear regime, while in the exponential model, the instability decays continuously. 
%%% After a strong linear growth, --> The strongest perturbation occurs at the linear growth stage.
Figure \ref{figturb} plots the evolution of the turbulence, 
measured as integrating the angular component of the kinetic energy density over the volume of shocked gas;
%measured as the angular component of the kinetic energy density,
%integrated over the volume of shocked gas: 
 \begin{equation}\chi={\int\rho u_{\theta}^{2} d\tau  \over  \int\rho d\tau}.\end{equation}
 After a strong linear growth, 
 the parameters converge to similarly small values of $\sim 0.2$ at $t'=2$ and $\sim 0.02$ at $t'=8$.
%%% I DON:T LIKE THIS SENTENCE Although higher compression ratios induce longer R-T fingers initially,
Since the ratio of the forward shock radius to the contact discontinuity radius increases with time, 
the R-T fingers eventually drop behind the forward shock, and instabilities fade.  
If the power-law model is extended by including a flat component to the inner ejecta and if the system evolves long enough such that the reverse shock begins to turn over to the stellar center (Fig. 1 of WC02), then the ratio of the forward shock radius to the contact discontinuity radius (which value remains fixed in the self-similar phase) starts to increase, and so the R-T convection declines as well.

\section{Discussion and Conclusions}

%%The energy losses resulted from the use of a low $\gamma$ 
%%through X-ray emission are small so that the dynamics of the flow are not affected by gas cooling

The acceleration by young SNR shocks of cosmic rays with high efficiency is expected to produce compression ratios of shocks that is considerably more than 4.
 To examine instability in a real situation in which the shocked gas is cooled by the escape of superthermal cosmic-ray particles from the intershock region, low adiabatic indices are input to the exponential density model, which is applicable to Type Ia SNRs. Since the extent of R-T mixing is determined by the deceleration and the density profile of the intershock region, such a simple model should be able to represent the overall dynamics of the R-T instability that develops in a typical Type Ia SNR.

When the adiabatic index is reduced, the SNR intershock gas is compressed toward the contact discontinuity, causing SN ejecta material to move closer to, or even slightly ahead of, the nominal shock radius. However, as the ratio of the forward shock radius to the contact discontinuity radius increases with time, the intershock structure becomes less compressed, and the R-T instabilities gradually decline.  Protrusion from the shocks (which can be caused by a limited grid resolution) is expected only in the linear growth stage of the R-T instability, probably soon after the SN explodes. In later dynamic phases of a remnant, Kelvin-Helmholtz instability overpowers Rayleigh-Taylor instability, and turbulence appearing in vortices dominates throughout the rest of the evolution, leaving the intershock flow rather featureless.

When the shock compression reaches $\sim 21$ ($\gamma$=1.1), the intershock region contracts to the extent that the reverse shock fronts resides at $\ge 87\%$ of the forward shock radius in the one-dimensional unperturbed case, or $\sim 77\%$ in the two-dimensional angle-averaged profile. However, a much smaller radius fraction of $\sim 2/3$ is observed for Tycho's SNR, which is consistent with adiabatic expansion. Contrary to previous thought, the compression does not alter the degree of convective mixing; strong perturbations rapidly stall in the linear instability growth stage. For a remnant near Tycho's present epoch, the mixing takes place
within a region below $\sim 87\%$ of the blast wave radius, independently of the value of $\gamma$. This result substantiates BE01's finding that the saturated instability growth rates in the power-law model do not detectably vary with $\gamma$.

The changes in the dynamics of the convective flow that are caused by the energy losses for low values of $\gamma$ are small when $\gamma \ge 1.2$. The extreme case of $\gamma =1.1$ involves significant, and probably not physically reasonable, losses of energy from the shock to cosmic rays. The escape of cosmic rays from shocks (by Alfven heating, among other processes) is not well understood, and detailed models of cosmic-ray-modified shock remain theoretical. In Ferrand {\it et al.} (2010), the kinetic model of Blasi (2004), which solves the particle distribution and the fluid velocity as functions of the particle momentum, is used to calculate the time-evolving shock properties and particle spectrum; a compact shock structure is revealed, but the R-T mixing remains not significantly affected. 
A magneto-hydrodynamic treatment of cosmic rays given in Skilling (1972 \& 1975), which includes
the particles as a second fluid component in a plasma system, can be utilized for future modeling of cosmic-ray modified SNRs (for applying such governing equations to Parker instabilities please see Lo, Ko, \& Wang 2011).

%Skilling (1972; 1975), in which the cosmic rays and thermal plasma are treated as two fluid components of a plasma system and 
%the interaction of cosmic-ray particles with the gas are represented by the cosmic-ray pressure $P_c$ and its gradient 
%(Lo, Ko, \& Wang 2011). 

%
%In the second short paper a separate module solves for the particle distribution f and the fluid velocity U,
%as functions of the particle momentum p. U represents the average fluid velocity felt by the particle
%while diffusing; since the diffusion depends mainly on p, so the U (this is already in Blasi 02).
%The new module in the code has input the velocity, compression, with as parameters injection rate
%and diffusion coefficient, and as output the new properties of the shock and the particle spectrum.
%

This study supports our earlier determination that enhanced R-T mixing is not responsible for the filaments and bumps seen on the outlines of young SNRs like Tycho's SNR.
Notably, the above result also holds for core-collapse SNRs, as determined using the power-law model after $t'\ge 0.7$, corresponding to $\sim 173$ yr for $M_{1.4}=1$ and $E_{51}=1$, or $\sim 890$ yr for a ten-fold explosion mass, when the reverse shock begins to enter the innermost plateau of the ejecta, terminating the self-similar expansion phase (WC02).       
The conclusion is based on the dynamics of two-dimensional flows. Three-dimensional simulations in BE01
reveal more deformation of the forward shock, with an overall drop in the average compression ratio and an increase in the width of the intershock region, but the convection remains relatively unaffected by the dimensionality of the simulation.

The contraction of the intershock structure depends on the pressure inside.
 External factors may adversely influence the flow.  For example, if a strong magnetic field is present in the ISM, then the pressure from the shocked toroidal magnetic field
 will distend the forward-shocked layer, preventing the forward shocked layer from shrinking, and further impeding the R-T mixing. 
Radio polarization measurements of several supernova remnants however show the dominance of the radial magnetic field  
(Dickel, ven Breugel, \& Milne 1991 for Tycho's SNR; Dickel, Strom, \& Milne 2001 for G315.4-23).
These observations also reveal 
the presence of out-moving Mach cones 
that resembles the features in the Vela SNR 
(Strom {\it et al.} 1995; 
Aschenbach, Egger, \& Tru\"umper 1995). 
%As well as Jun & Norman (1996b), there should be a reference to Jun, Jones & Norman (1996, ApJ, 468, L59). These two papers suggest a slightly different scenario of how ejecta clumps could approach the blast wave. A forward shock running into a clumpy medium advects vorticity post shock which enhances the R-T instability. It remains to be seen whether this actually happens in reality, though people are starting to think about  shock propagation into turbulent media. In any case, I think this pair of references and the model they discuss should be given a bit more prominence in the paper.
On the other hand, if the forward shock encounters a clumpy interstellar medium,
then the vortices that are generated behind the shock front can channel the R-T instability to enhance the mixing and create protrusions in the supernova remnant shell (Jun, Jones, \& Norman 1996). Fujita {\it et al.} (2009) suggested that particles can be accelerated close to molecular clouds even after the SNR becomes radiative. However, the process for the Rayleigh-Taylor fingers reach the forward shock (depending on the properties and distribution of the clouds that are engulfed by the supernova shock) is transient, and whether the shock-cloud interaction can overcome the global decay of instability in a remnant like Tycho’s and in a typical, turbulent ISM is unclear.

The possibility that the mixture of heavy-element ejecta near the blast wave is produced by R-T convection during the ejecta-dominated stage of an SNR is excluded, and we speculate that the injection of cosmic-ray particles at Tycho's forward shock is not sufficiently strong to alter the shock properties, including the compression ratio. 
The extreme case of $\gamma=1.1$ suggests that the outline is not to be substantially perturbed. 
Even if the injection and efficient acceleration of cosmic rays at the shock suffice, 
%Even with very efficient acceleration of cosmic rays at the shock,
the extent of mixing is not expected to be significantly enhanced relative to the adiabatic case.

While our result contradicts the interpretation of Warren {\it et al.} (2005) based on X-ray data, the simulations 
simply reflect the basic dynamic properties of flow in multi dimensions that X-ray power spectra may not adequately reveal.
The increased mixing is best explained by
 the interaction clumpy ejecta that is produced by the nickel bubble effect with the intershock region of the remnant.  
In this scenario, the radioactive heating of $^{56}Ni$ inflates the central ejecta and drives a strong thin shock to compress the outer heavy elements into dense ejecta clumps.
Since the heavy elements were shocked before they underwent reverse shock,
 this process is likely to facilitate the creation of more energetic particles, as well as denser clumping, in a supernova remnant.

\section*{Acknowledgments}
%Acknowledgments

The author would like to thank Roger Chevalier, John Blondin, Vikram Dwarkadas,
for correspondence and discussions. Special thanks are also due to Federico Fraschetti and Glenn Allen for useful correspondence on hydrodynamic simulations and observations of cosmic rays in SNRs. The author thanks the referee Martin Laming for his helpful comments on the manuscript.
This work was supported by
the National Science Council of Taiwan 
%for financially supporting this research 
and the National Center for High-performance Computing of Taiwan.
%for providing computing time and facilities.
%Ted Knoy is appreciated for his editorial assistance.

%I thank Professor N. Kameswara Rao for some helpful suggestions,
%Dr H. C. Bhatt for a critical reading of the original version of the
%paper and an anonymous referee for very useful comments that improved
%the presentation of the paper.

%\begin{thebibliography}{99}

\end{document}